\documentclass[prl,twocolumn,superscriptaddress,showpacs,floatfix]{revtex4}
\usepackage{graphicx,amsmath,amssymb,bm}
\usepackage{amsfonts}
\usepackage{braket}

\usepackage{graphicx}
\usepackage{dcolumn}
\usepackage{bm}

\newcommand*{\intsum}{\mathop{\ooalign{\raise0.2pt\hbox{$\int$}%
            \cr\lower0.3pt\hbox{$\Sigma$}}}}

\begin{document}

\preprint{ }

\title{Elastic proton scattering of medium mass nuclei from coupled-cluster theory}

\author{G.~Hagen}
\affiliation{Physics Division, Oak Ridge National Laboratory,
Oak Ridge, TN 37831, USA}
\affiliation{Department of Physics and Astronomy, University of
Tennessee, Knoxville, TN 37996, USA}
\author
{N. Michel}
\affiliation{Department of Physics and Astronomy, University of Tennessee, Knoxville, TN 37996, USA}

\date{\today}

\begin{abstract}
  Using coupled-cluster theory and interactions from chiral effective
  field theory, we compute overlap functions for transfer and
  scattering of low-energy protons on the target nucleus $^{40}$Ca.
  Effects of three-nucleon forces are included phenomenologically as
  in-medium two-nucleon interactions.  Using known asymptotic forms
  for one-nucleon overlap functions we derive a simple and intuitive
  way of computing scattering observables such as elastic scattering
  phase shifts and cross sections.  As a first application and
  proof-of-principle, we compute phase shifts and differential
  interaction cross sections at energies $9.6$~MeV and $12.44$~MeV and
  compare with experimental data. Our computed diffraction minima are
  in good agreement with experiment, while we tend to overestimate the
  cross sections at large scattering angles.
\end{abstract}


\maketitle

\emph{Introduction} -- With the advent of accelerators of new
generation, providing radioactive ions beams, it becomes possible to
synthesize nuclei far from the valley of stability. Little is known
about these nuclei.  For example, the nuclear interaction is not well
understood at large proton-to-neutron ratio, and it appears that its
effects are radically different close to drip-lines compared to the
vicinity of the valley of stability. The most striking example is the
redistribution of shell closures, which have been noticed to be
different from the usual magic numbers present in well-bound nuclei
\cite{hagen2012b,shell_closure_drip_lines_exp}. In order to precisely
study nuclei close to drip-lines, new theoretical tools must be
developed, going further from the standard structure methods of
nuclear analysis, based on the use of standard shell model, treating
all nuclear states as well bound, and, in reaction theory, of optical
potentials fitted from experimental data. Indeed, both nuclear
structure and reactions must be unified, so that both inter nucleon
correlations and scattering degrees of freedom are included within the
same framework \cite{jacek2007}. Microscopic models aiming at a
unification of nuclear structure and reactions have made great
progress over the last decade. The Resonating Group Method within the
No Core Shell Model, has successfully described nucleon and deuteron
scattering and fusion in light nuclei \cite{RGM_NCSM}, the
Green's Function Monte Carlo method has been used to describe elastic
scattering on ${^4}$He and in the computation of asymptotic
normalization coefficients in light nuclei \cite{GFMC_scat}, and
finally the Self Consistent Green's Function (SCGF) method has been
applied to the microscopic calculation of optical potentials and
proton scattering on $^{16}$O \cite{SCGF_O16}.

An interesting avenue to compute nuclear reactions microscopically
lies in coupled-cluster theory. The coupled-cluster method is a
microscopic theory which comes at a relatively low computational cost
and at the same time can provide accurate description of low-lying
states and properties of nuclei with closed (sub-)shells
\cite{ccm,bartlett07}. The coupled-cluster method has in the last
decade made significant progress in computing structure of nuclei from
the valley of stability towards the neutron dripline. Using a Berggren
basis \cite{berggren1968} consisting of bound, resonant and scattering
states, both loosely bound and unbound states have been accurately
computed within the coupled-cluster formalism \cite{hagen2007b,
  hagen2010a}. However, so far no attempt has been made to apply the
coupled-cluster method to compute reaction observables, and it is the
aim of this work to fill this gap and to develop a new formalism to
compute reaction observables such as elastic scattering cross sections
using micrcoscopic coupled-cluster theory.  As a first application we
will consider the elastic scattering reaction $^{40}$Ca(p,p)$^{40}$Ca,
whose phase shifts and differential elastic cross sections will be
evaluated at low energies.

\emph{Hamiltonian and treatment of the infinite-range Coulomb
  interaction} -- The intrinsic $A-$nucleon Hamiltonian consists of
kinetic, nuclear and Coulomb parts,
\begin{equation}
\label{ham}
\hat{H} = \sum_{1\le i<j\le A}\left({(\vec{p}_i-\vec{p}_j)^2\over 2mA} + \hat{V}
_{NN}^{(i,j)} + \hat{V}_{\rm Coul}^{(i,j)} + \hat{V}_{\rm 3N eff}^{(i,j)}\right).
\end{equation}
Here, the intrinsic kinetic energy depends on the mass number $A$.
The potential $\hat{V}_{NN}$ denotes the chiral $NN$ interaction at
next-to-next-to-next-to leading order~\cite{entem2003,machleidt2011}
(with cutoff $\Lambda=500$ MeV), $\hat{V}_{\rm Coul}$ is the Coulomb
interaction, while $\hat{V}_{\rm 3N eff}$ is a schematic potential
based on the in-medium chiral $NN$ interaction by Holt~{\it et
  al.}~\cite{holt2009}. The potential $\hat{V}_{\rm 3N eff}$ results
from integrating one nucleon in the leading-order chiral 3NF over the
Fermi sphere with Fermi momentum $k_F$ in symmetric nuclear matter. It
depends formally on the Fermi momentum $k_F$, the low-energy constants
$c_D$ and $c_E$ of the short-ranged contributions to the leading-order
chiral 3NF, and the chiral cutoff. The latter is equal to the value
employed in the chiral $NN$ interaction~\cite{entem2003}. In this work
we employ the parameters $k_F=0.95$~fm$^{-1}$, $c_D=-0.2$ and
$c_E=0.735$ which was recently applied for the study of shell
evolution in neutron rich calcium isotopes~\cite{hagen2012b}.

Let us briefly discuss our treatment of the short and long-range parts
of the Hamiltonian in Eq.~(\ref{ham}). The nuclear interaction
$\hat{V}_{NN}$ is of short range and is adequately expanded in a basis
of harmonic oscillator states (see Ref.~\cite{PRC_HO_expansion} for
details). The difficulty induced by the infinite-range character of
the Hamiltonian is thus embodied in the Coulomb interaction
$\hat{V}_{\rm Coul}$, asymptotically behaving as $(Z-1) e^2 / r$, with
$r$ the distance between the isolated proton and the center of charge
of the remaining part of the nucleus. Clearly, it is insufficient to
treat $\hat{V}_{Coul}$ with a harmonic oscillator expansion as we do
for $\hat{V}_{NN}$. A solution to this problem has been formulated in
Ref.~\cite{PRC_isospin_symmetry_breaking}.  For this, the Coulomb
interaction is rewritten as sum of two terms:
\begin{eqnarray}
V_{\rm Coul} = U_{\rm Coul}(r) + [V_{\rm Coul} - U_{\rm Coul}(r)] \label{Coulomb_separation},
\end{eqnarray}
where one demands the Coulomb one-body potential $U_{\rm Coul}(r)$ to
behave as $(Z-1) e^2 / r$ for $r \rightarrow +\infty$. In this work we
choose $U_{\rm Coul}(r) = {\rm erf}(\alpha r) (Z-1) e^2 / r$, where
${\rm erf}$ is the error function and $\alpha = \pi/4
~\mathrm{fm}^{-1}$.  Thus, the $[V_{\rm Coul} - U_{\rm Coul}(r)]$ term
is short-ranged, so that one can use the harmonic oscillator expansion
method of Ref.~\cite{PRC_HO_expansion} to calculate its matrix
elements.  Note that the $r$ coordinate can be taken with respect to
the origin of the laboratory, because center of charge effects are
negligible in the asymptotic region on the one hand, and for a medium
mass nucleus such as $^{40}$Ca on the other hand.

In order to account for the scattering continuum using the
coupled-cluster formalism, it is convenient to express the Hamiltonian
for given partial waves in a basis of spherical Bessel
functions~\cite{hagen2010a}.  Thus, in order to proceed, we express
$U_{\rm Coul}(r)$ in momentum space, and write it in the following
way,
\begin{eqnarray}
  \nonumber
  U_{\rm Coul}(k,k') =  \langle k | U_{\rm Coul}(r) - \frac{(Z-1) e^2}{r} | k' \rangle + \\ 
  \frac{(Z-1) e^2}{\pi} Q_\ell \left( \frac{k^2 + {k'}^2}{2 k k'} \right)  \label{Uc_k_kp},
\end{eqnarray}
where $\ell$ is the orbital angular momentum of the considered partial
wave and $Q_\ell$ is the Legendre function of the second kind
\cite{Abramowitz_Stegun}. As the the first term of Eq.~(\ref{Uc_k_kp})
decreases very quickly for $r \rightarrow +\infty$, it can be
calculated by numerical integration. However, the second term presents
a logarithmic singularity at $k = k'$.  In order to counter this state
of affairs, we will follow the off-diagonal method introduced in
Ref.~\cite{PRC_Coulomb_Berggren}. It consists in replacing the infinite
value $Q_\ell(1)$ in Eq.~(\ref{Uc_k_kp}) occurring at $k = k'$ by a
finite value depending on the discretization used (see
Ref.~\cite{PRC_Coulomb_Berggren} for method and details).

\begin{figure}[thbp]
  \begin{center}
    \includegraphics[width=0.5\textwidth, clip]{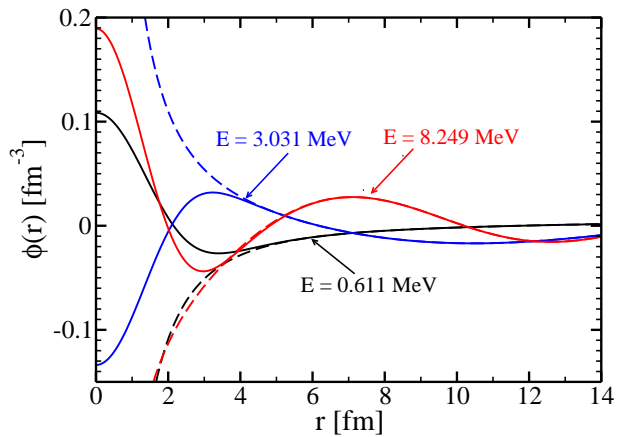}
  \end{center}
  \caption{(Color online) Scattering $s$-wave functions $\varphi(r)$
    obtained from diagonalization of the one-body Hamiltonian defined
    in Eq.~(\ref{one_body_h}) in momentum space (solid lines) and their
    asymptotic expansion defined in Eq.~(\ref{asymptotic_Coulomb_fit})
    (dashed lines) as a function of radius, provided in fm. Wave
    functions are given in units of fm$^{-3}$ and their energy is
    written on the figure in units of MeV}
  \label{scat_Coulomb_fit}
\end{figure}

In order to show the precision of the method in the context of
momentum space calculation, we will diagonalize with a basis of Bessel
functions the one-body Hamiltonian for the $\ell = 0$ partial wave
studied in Ref.~\cite{PRC_Coulomb_Berggren}, which reads:
\begin{eqnarray}
h = \frac{\hat{p}^2}{2m} - V_o \left[ 1 + \exp \left( \frac{r-R_0}{d} \right) \right]^{-1} + U_{\rm Coul}(r) \label{one_body_h}
\end{eqnarray}
where $m$ is the proton mass, $d$ = 0.65 fm, $R_0$ = 3 fm, $V_o$ = 52
MeV, and $U_{\rm Coul}(r)$ is the Coulomb potential of
\cite{PRC_HO_expansion}.  Obtained scattering wave functions have been
fitted for large $r$ with their asymptotic limit equal to:
\begin{eqnarray}
C_F \frac{F(\ell,\eta,k r)}{r} + C_G \frac{G(\ell,\eta,k r)}{r} \label{asymptotic_Coulomb_fit}
\end{eqnarray}
where $F(\ell,\eta,x)$ and $G(\ell,\eta,x)$ are respectively the
regular and irregular Coulomb wave functions \cite{Abramowitz_Stegun},
$\eta$ is the Sommerfeld parameter and $C_F$ and $C_G$ are integration
constants. Regular and irregular Coulomb wave functions are evaluated
numerically using the publicly available {\sc cwfcomplex} code
\cite{Coulomb_code_CPC}, while $C_F$ and $C_G$ constants are
determined by fitting Eq.~(\ref{asymptotic_Coulomb_fit}) to the
considered scattering wave functions at $r = 10$ fm. Results are
depicted in Fig.~\ref{scat_Coulomb_fit}. It is therein clear that
their asymptotic behavior is very well reproduced, as Coulomb
asymptotic expansions and diagonalized scattering wave functions are
virtually indistinguishable for $r > 7$ fm. This proves that the
infinite-range character of the Coulomb interaction can be handled
precisely with Fourier-Bessel transform, so that reactions involving
protons, such as elastic scattering, can be undertaken.

\emph{One-nucleon overlap functions and coupled-cluster theory} -- The
scattering of a nucleon on a target $A$ can be described by the one
nucleon overlap function. The one-nucleon radial overlap function
$O_{A}^{A+1}(lj;r)$ is defined microscopically as the overlap between
two independent many-nucleon wave functions of $A$ and $A+1$ nucleons,
\begin{equation}
O_{A}^{A+1}(lj;kr) = \intsum_n\Braket{A+1||\tilde{a}_{nlj}^{\dagger}||A}\phi_{nlj}(r).
\label{equDefOverlap}
\end{equation}
The double bar denotes a reduced matrix element, and the integral-sum
over $n$ represents both the sum over the discrete spectrum and an
integral over the corresponding continuum part of the spectrum. The
creation operator $\tilde{a}_{nlj}^\dagger$ is a spherical tensor of
rank $j$. The radial single-particle basis function is given by the
term $\phi_{nlj}(r)$, where $l$ and $j$ denote the single-particle
orbital and angular momentum, respectively, and $n$ is the nodal
quantum number. The isospin quantum number has been suppressed.  The
one-nucleon overlap functions describes the capture or scattering of
an incoming particle with quantum numbers $lj$ on the target nucleus
$A$ and with the final state $A+1$ being either a bound or a
scattering state. The momentum $k$ is given from the energy difference
$k = \sqrt{2\tilde{m} (E^{A+1} - E^A)}/\hbar$, in the case of $A$ and $A+1$
being in their ground state $E^{A+1}-E^A$ is the one-nucleon separation
energy in the $A+1$ nucleus (here $\tilde{m}=(1-1/A)m$).  We emphasize
that the overlap function is defined microscopically and independently
of the single-particle basis. It is uniquely determined by the
many-body wave functions $\ket{A}$ and $\ket{A+1}$. $\ket{A}$ and
$\ket{A+1}$ can in general either be in their ground- or any excited
state. However, in this work we are interested in low-energy elastic
scattering, which implies that the target nuclues $\ket{A}$ is in its
ground state before and after the scattering. The one-nucleon overlap
functions are formally solutions of the Dyson equation, which can be
written in a Schr{\"o}dinger like form where the self-energy takes the
place of a non-local and energy dependent optical potential
\cite{SCGF_O16}.  Outside the range $R$ of the optical potential, the
one-nucleon overlap functions for bound $A+1$ states takes ($k =
i\kappa$) the form,
\begin{equation} 
O_{A}^{A+1}(lj;kr) =  C_{lj}(i\kappa) { W_{-\eta, l+1/2}(i\kappa r)\over r}
\label{eq:bound}
\end{equation}
and for $A+1$ scattering states ($k>0$),  
\begin{equation}
O_{A}^{A+1}(lj;kr)  =   B_{lj}(k)\left[ F_{l,\eta}(kr) - \tan\delta_l (k) G_{l,\eta}(kr) \right]. 
\label{eq:scatt}
\end{equation} 
Here $W_{-\eta, l+1/2}$ is the Whittaker function, $F_{l,\eta} $ and
$G_{l,\eta}$ the regular and irregular Coulomb wave functions, $\eta$
is the Sommerfeld parameter ($\eta = (Z-1)e^2 \sqrt{ \tilde{m}/2|E|}
$), $ C_{lj}(i\kappa)$ is the asymptotic normalization coefficient
(ANC), $\tan\delta_l (k)$ is the $l$'th partial wave scattering phase
shift at momentum $k$, and $ B_{lj}(k)$ is an arbitrary normalization
constant for the scattering states. In order to compute the phase
shifts at a given energy, it is sufficient to know the one-nucleon
overlap function $O_{A}^{A+1}(lj;kr)$ and it's derivative at a given
radius $r>R$. In order to obtain $O_{A}^{A+1}(lj;kr)$ we need to solve
for the ground state of the target nucleus $A$ and the ground- and
excited scattering states in the residual nucleus $A+1$. The
coupled-cluster method is a very efficient tool for the computation of
ground- and low-lying excited states in nuclei with a closed
(sub-)shell structure and their neighbors.  In this work the target
nucleus $A$ is a closed shell nucleus, and we use the coupled-cluster
method to compute the ground state of $A$, i.e.  $ \vert A \rangle =
e^T\vert \phi_A\rangle$. Here $\vert \phi_A \rangle $ is the
Hartree-Fock reference state while $T$ is a linear combination of
particle-hole excitation operators.  For the residual $A+1$ nucleus we
use particle-attached equation-of-motion coupled cluster theory to
obtain the ground- and excited states, and the $A+1$ wave functions
are therefore given by $ \langle A+1\vert_\mu = \langle \phi_A\vert
L_\mu^{A+1}e^{-T}$, with $ L_\mu^{A+1} $ a linear combination of
one-particle, and two-particle-one-hole excitations operators (details
on our implementation are presented in
Refs.~\cite{hagen2008,hagen2010b}). Inserting these expressions for
the $A$ and $A+1$ systems into Eq.~(\ref{equDefOverlap}), we obtain the
coupled-cluster formulation of the one-nucleon overlap functions,
\begin{equation}
O_{A}^{A+1}(lj;kr) = \intsum_n\Braket{\phi_A||L_\mu^{A+1}\overline{ \tilde{a}_{nlj}^{\dagger} }||\phi_A}\phi_{nlj}(r),
\label{equDefOverlap2}
\end{equation}
here $\overline{\tilde{a}_{nlj}^{\dagger}} =
e^{-T}\tilde{a}_{nlj}^{\dagger}e^T $ is the similarity transformed
creation operator. The derivation of the diagrammatic and algebraic
expressions of Eq.~(\ref{equDefOverlap2}) and
$\overline{\tilde{a}_{nlj}^{\dagger}} $ can be found in
Ref.~\cite{jensen2010}. Note that in order to compute the radial
overlap in Eq.~(\ref{equDefOverlap2}) we need to use the same mass
number $(41)$ in the intrinsic kinetic energy of the $A$ and $A+1$
Hamiltonians in Eq.~(\ref{ham}). This introduces a small error in the
ground state of the target nucleus $A$.  However, this error decreases
rapidly with increasing mass, and we estimate that the error is of the
order of $100-200$~keV in the relative energy entering the overlap
function \cite{hagen2012a}.

In our coupled-cluster calculations we use a
model space consisting of $N_{\rm max}=17$ major spherical oscillator
shells with the oscillator frequency $\hbar\omega = 26$~MeV. This is a
sufficently large model space to reach practically converged results
for the ground state of $^{40}$Ca (see Ref.~\cite{hagen2012b}). In
order to properly account for scattering continuum in $^{41}$Sc we use
a Gamow-Hartree-Fock basis~\cite{michel2009} for the relevant
proton partial waves. In constructing the single-particle basis with
the correct treatment of long-range Coulomb effects, we use the
off-diagonal method in momentum space and discretize the one-body
momentum space Schr{\"o}dinger equation with 50 mesh points. We find
that this is a sufficiently large number of mesh points in order to
obtain the correct Coulomb asymptotics necessary to describe proton 
elastic scattering on $^{40}$Ca.

\begin{figure}[thbp]
  \begin{center}
    \includegraphics[width=0.5\textwidth, clip]{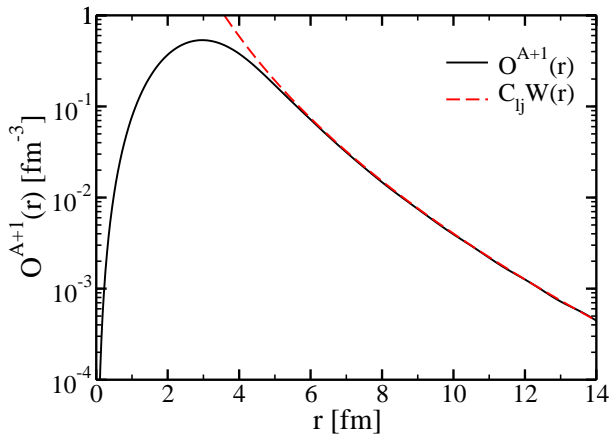}
  \end{center}
  \caption{(Color online) Radial overlap function $O_{A}^{A+1}(lj;kr)$ between the 
    ground state of $^{40}$Ca and the $J^\pi= 7/2^-$ ground state of $^{41}$Sc (solid line), also shown is the 
    corresponding Whittaker function  $C_{lj}W_{-\eta, l+1/2}(i\kappa r)/r $ for the $f_{7/2}$ proton
    partial wave (dashed line).}
  \label{bound_Coulomb_fit}
\end{figure}

\begin{figure}[thbp]
  \begin{center}
    \includegraphics[width=0.5\textwidth, clip]{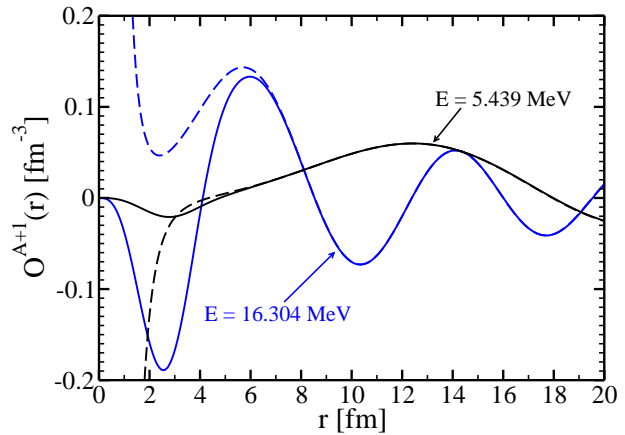}
  \end{center}
  \caption{(Color online) Radial overlap functions
    $O_{A}^{A+1}(lj;kr)$ between the ground state of $^{40}$Ca and two
    $J^\pi= 7/2^-$ scattering states in $^{41}$Sc (solid lines), also
    shown are the corresponding Coulomb scattering functions $
    B_{lj}(k)\left[ F_{l,\eta}(kr) - \tan\delta_l (k) G_{l,\eta}(kr)
    \right] $ (dashed lines).}
  \label{scat_overlap_functions}
\end{figure}

\begin{figure}[thbp]
  \begin{center}
    \includegraphics[width=0.5\textwidth, clip]{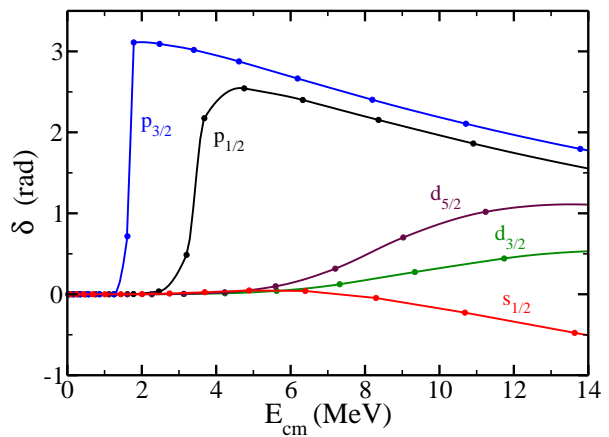}
  \end{center}
  \caption{(Color online)
    Computed phase-shifts for elastic proton scattering on $^{40}$Ca for
    low-lying partial waves and energies below $14$~MeV.}
  \label{phase_shifts}
\end{figure}

\begin{figure}[thbp]
  \begin{center}
    \includegraphics[width=0.5\textwidth, clip]{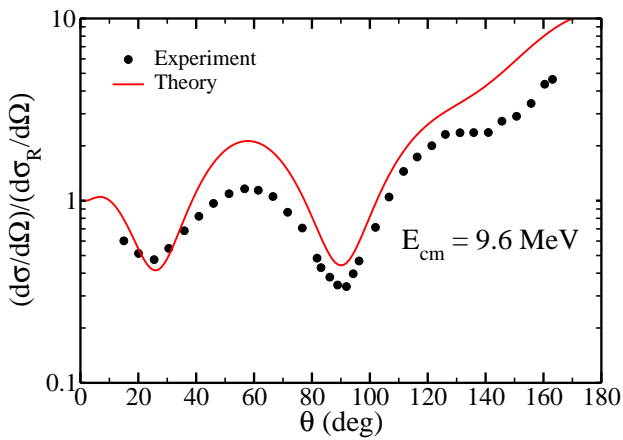}
  \end{center}
  \caption{(Color online) Differential cross section divided by Rutherford cross section 
    for elastic proton scattering on $^{40}$Ca at $E_{\rm cm}=9.6$~MeV (solid line), 
    experimental data (dots) are taken from ~Ref.\cite{oers1971}.}
  \label{diff_cross_section_9_6}
\end{figure}

\begin{figure}[thbp]
  \begin{center}
    \includegraphics[width=0.5\textwidth, clip]{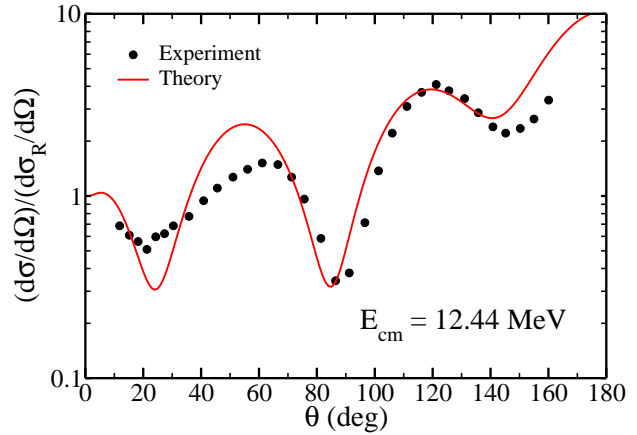}
  \end{center}
  \caption{(Color online) Differential cross section divided by Rutherford cross section 
    for elastic proton scattering on $^{40}$Ca at $E_{\rm cm}=12.44$~MeV(solid line), 
    experimental data (dots) is taken from ~Ref.\cite{oers1971}.}
    \label{diff_cross_section_12_44}
\end{figure}

\emph{Results} -- Figure~\ref{bound_Coulomb_fit} shows the computed
radial overlap function for the ground state of $^{40}$Ca with the
$J^\pi= 7/2^-$ ground state of $^{41}$Sc on a logarithmic scale. Our
computed proton separation energy for $^{41}$Sc is $S^{\rm CC}_{\rm p}
= 0.71$~MeV, which is in good agreement with the experimental proton
separation energy $S^{\rm Exp}_{\rm p} = 1.09$~MeV. From the radial
overlap function and the separation energy we can compute the
behaviour of the the overlap function at distances beyond the range of
the nuclear interaction according to Eq.~(\ref{eq:bound}). It is clearly
seen that the overlap function and the known asymptotic form
completely overlap for distances larger than $r \sim 8$~fm. 
Figure~\ref{scat_overlap_functions} shows the computed radial overlap
functions for the ground state of $^{40}$Ca with two $J^\pi= 7/2^-$
scattering states of $^{41}$Sc, at the energies $E=5.439$~MeV and $E=
16.304$~MeV, respectively. As we found for the bound overlap function
shown in Fig.~\ref{bound_Coulomb_fit}, we see that the radial overlap
function for scattering states and the known asymptotic forms
completely overlap for distances larger than $r \sim 8$~fm.
By matching the asymptotic forms of the overlap functions given in
Eq.~(\ref{eq:scatt}) with the computed overlap functions, it is clear
that we can determine the corresponding elastic scattering phase-shift
at the computed scattering energy. Figure~\ref{phase_shifts} shows our
computed scattering phase shifts for proton elastic scattering on
$^{40}$Ca for the $s_{1/2}, p_{1/2}, p_{3/2}, d_{3/2}, d_{5/2}$
partial waves at energies below $14$~MeV. The solid dots correspond to
the computed scattering energies, and we used cubic spline to
interpolate between the discrete set of scattering energies.  We
clearly see the appearance of a narrow resonance in the $p_{3/2}$
partial wave around $\sim 1.6$~MeV, while a broader resonance appear
in the $p_{1/2}$ partial wave at around $\sim 3.4$~MeV. In order to
check our results we computed the low-lying resonances in $^{41}$Sc
using a complex Gamow-Hartree-Fock basis \cite{hagen2010a}, and we
found a $J^\pi= 3/2^-$ resonance at the energy $E = 1.61 - 0.001i
$~MeV and a $J^\pi= 1/2^-$ resonance at the energy $E = 3.42 - 0.20i
$~MeV. Clearly these energies are consistent with the resonances
appearing in the $p_{3/2}$ and $p_{1/2}$ elastic scattering
phase-shifts in Fig.~\ref{phase_shifts}.
From the scattering phase-shifts we can compute the differential cross
section for elastic proton scattering as described in
e.g. Ref.~\cite{fresco}. Figures~\ref{diff_cross_section_9_6} and
\ref{diff_cross_section_12_44} show the differential cross section
divided by the Rutherford cross section for elastic proton scattering
on $^{40}$Ca at the relative-center-of mass energies $E_{\rm
  cm}=9.6$~MeV and $E_{\rm cm}=12.44$~MeV, respectively. All partial
waves for $l\leq 2$, were included in the computation of the cross
sections. Overall we get good agreement between our calculated cross
sections and the experimental cross sections. In particular we see
that our computed minima are in good agreement with the experimental
minima, while we tend to overestimate the cross sections at large
scattering angles.  The overestimated cross sections at large angles
is most likely due to the fact that we do not account for intermediate
excitations that takes place above the deuteron threshold, and these
excitations generally cause absorption and reduces the cross section
at large angles. Going beyond $2p$-$1h$ excitations for the
computation of the $A+1$ wave functions will account for such effects,
and we are working towards such improvements in our approach. We also
computed the cross sections including the $f_{5/2},f_{7/2}, g_{7/2},
g_{9/2}$ partial waves, however the agreement with data did not
improve. This is most likely due to the fact that our computations for
the $ f_{5/2} $ and $ g_{9/2} $ partial waves finds very narrow
resonances at too low energy energies as compared to experiment.  This
is also consistent with the $\sim 10$ MeV overbinding we get for
$^{40}$Ca using the Hamiltonian in Eq.~(\ref{ham}) \cite{hagen2012b}.

\emph{Conclusions} -- Using coupled-cluster theory, we computed cross
sections for elastic scattering of protons on $^{40}$Ca, at the
center-of-mass energies $9.6$~MeV and $12.44$~MeV, respectively. We
found good agreement for our computed diffraction minima with
experiment, while we tend to overestimate the cross sections at large
scattering angles.  The key ingredients for computing observables for
proton scattering are; (i) the one-nucleon overlap function computed
from microscopic coupled-cluster theory, and (ii) a single-particle
basis that has the correct Coulomb asymptotics. We showed that the
newly developed \textit{off-diagonal} method is a very accurate method
to compute Coulomb scattering wave functions in momentum space. The
fast convergence of the scattering wave functions with increasing
number of mesh-points makes this basis an ideal starting point for
computing reaction observables.  This work constitute the first
successful application of coupled-cluster theory to nuclear reactions,
and we believe it makes a significant leap forward in linking
reactions with microscopic structure calculations.

\begin{acknowledgments}
  We acknowledge valuable discussions with T. Papenbrock and F. Nunes.
  This work was supported by the Office of Nuclear Physics,
  U.S.~Department of Energy (Oak Ridge National Laboratory).  This
  work was supported in part by the U.S. Department of Energy under
  Grant Nos.~DE-FG02-03ER41270 (University of Idaho),
  DE-FG02-96ER40963 (University of Tennessee), and DE-FC02-07ER41457
  (UNEDF SciDAC).  This research used computational resources of the
  National Center for Computational Sciences, the National Institute
  for Computational Sciences.
\end{acknowledgments}

\end{document}